\documentclass[sigconf,10pt]{acmart}

\usepackage{amsmath}

\usepackage{diagbox}
\usepackage{slashbox}
\usepackage{amsfonts} 
\usepackage{xcolor} 
\usepackage{graphicx} 
\usepackage{comment} 
\usepackage{lipsum} 
\usepackage{balance}
\usepackage{enumitem}
\usepackage{caption}
\usepackage{subcaption}
\usepackage{multirow}
\usepackage{array}

\newcolumntype{L}[1]{>{\raggedright\let\newline\\\arraybackslash\hspace{0pt}}m{#1}}
\newcolumntype{C}[1]{>{\centering\let\newline\\\arraybackslash\hspace{0pt}}m{#1}}
\newcolumntype{R}[1]{>{\raggedleft\let\newline\\\arraybackslash\hspace{0pt}}m{#1}}

\newcommand{\coloroverride}[2]{\textcolor{#1}{#2}}

\newcommand{\chk}[1]{\coloroverride{red}{#1}} 

\graphicspath{{Figures/}}

\makeatother

\begin{document}

\copyrightyear{2024}
\acmYear{2024}
\setcopyright{acmlicensed}\acmConference[ACM MobiCom '24]{The 30th Annual International Conference on Mobile Computing and Networking}{November 18--22, 2024}{Washington D.C., DC, USA}
\acmBooktitle{The 30th Annual International Conference on Mobile Computing and Networking (ACM MobiCom '24), November 18--22, 2024, Washington D.C., DC, USA}
\acmDOI{10.1145/3636534.3697319}
\acmISBN{979-8-4007-0489-5/24/11}


 \title{Data Driven Environment Classification  Using Wireless  Signals    }




\author{Hossein Nasiri}
\affiliation{%
  \institution{Univ. of Notre Dame}
  \country{}
}
\email{hnasiri2@nd.edu}

\author{Seda Dogan-Tusha}
\affiliation{%
  \institution{Univ. of Notre Dame}
  \country{}
}
\email{stusha@nd.edu}

\author{M. I. Rochman}
\affiliation{%
  \institution{Univ. of Notre Dame}
  \country{}
}
\email{mrochman@nd.edu}

\author{Monisha Ghosh}
\affiliation{%
  \institution{Univ. of Notre Dame}
  \country{}
}
\email{mghosh3@nd.edu}

\begin{abstract}
Robust classification of the operational environment of wireless devices is becoming increasingly important for wireless network optimization, particularly in a shared spectrum environment. Distinguishing between indoor and outdoor devices can enhance reliability and improve coexistence with existing, outdoor, incumbents. For instance, the unlicensed but shared 6 GHz band (5.925 - 7.125 GHz) enables sharing by imposing
lower transmit power for indoor unlicensed devices and a spectrum coordination requirement for outdoor devices.
Further, indoor devices are prohibited from using battery power, external antennas, and weatherization to prevent outdoor operations.
As these rules may be circumvented, we propose a robust indoor/outdoor classification method by leveraging the fact that the radio-frequency environment faced by a device are quite different indoors and outdoors.
We first collect signal strength data from all cellular and Wi-Fi bands that can be received by a smartphone in various environments (indoor interior, indoor near windows, and outdoors), along with GPS accuracy, and then evaluate three machine learning (ML) methods: deep neural network (DNN), decision tree, and random forest to perform classification into these three categories. Our results indicate that the DNN model performs the best, particularly in minimizing the most important classification error, that of classifying outdoor devices as indoor interior devices.
\end{abstract}


\maketitle


\section{Introduction}
\subsection{Indoor/Outdoor Classification} \label{sec:imp_of_class}

The continuing growth of wireless applications requires access to new spectrum.  The mid-band frequency range (1-10 GHz) is industry's preference due to the balance between coverage and capacity compared to low-band (<1 GHz) and mmWave (>24 GHz)~\cite{FCC3}. However, the widespread deployment of various wireless technologies
has led to congestion in this spectrum band, which in turn highlights the need for spectrum sharing between multiple wireless networks, maximizing the efficiency of spectrum use~\cite{IoT,CAS}.
There have been efforts by regulatory bodies to promote spectrum sharing in the mid-band. Recently, the 6 GHz (5.925 - 7.125 GHz) band has been a particular focus of spectrum sharing with various incumbents, notably radio astronomy, fixed microwave links, cable television relay services (CTRS), satellite, and mobile Broadcast Auxiliary Services (BAS).

In the U.S., the Federal Communications Commission (FCC) has permitted unlicensed operation \cite{FCC1,FCC2}, with power limitations based on indoor/outdoor location of access points (APs) and clients. The FCC defines three power regimes: Standard Power (SP), with the highest power but subject to Automated Frequency Coordination (AFC); Low Power Indoor (LPI), with lower power but without AFC requirement, deployed indoors without battery-power, weatherization, and detachable antennas, and; Very Low Power (VLP), with the lowest power but without AFC or indoor requirements.
Moreover, all client devices are subject to 6 dB lower transmit power than the AP on all power regimes. While this is intended to mitigate outdoor interference, it arguably penalizes even indoor clients connected to LPI APs, as indoor clients are unlikely to cause significant interference~\cite{dogan2023evaluating,dogan2023indoor}. Accurately determining the indoor/outdoor location of clients could eliminate this 6 dB power reduction, allowing for more efficient use of the spectrum without compromising interference protection.

Meanwhile, the U.K. telecommunications regulator, Ofcom, has recently proposed innovative hybrid sharing approaches for the upper 6 GHz band (6.425 - 7.125 GHz), including an ``indoor/outdoor split'' model, \textit{i.e.}, the same frequency band for outdoor cellular use and indoor Wi-Fi use. However, the implementation of such strategies underscores the need for accurate Indoor/Outdoor (I/O) classification.

\vspace{-1em}

\subsection{Related Work}

Given the importance of the I/O classification, how can the environment be detected using just wireless data? Much like the discernible differences in visual aspects between indoor and outdoor spaces, Radio Frequency (RF) observations also exhibit distinguishable features that set apart indoor and outdoor environments. For instance, the Wi-Fi Received Signal Strength Indicator (RSSI) is typically higher in indoor environments, while the Reference Signal Received Power (RSRP) of cellular signals and GPS accuracy is higher outdoors. This is expected due to the deployment location of the transmitters of the respective networks~\cite{lee2017classification, app10020500}. ML methods can discern the inherent relationships between a dataset and the labeled environment, allowing them to generate precise predictions for new data. 

There are a variety of commonly available RF signals that are received indoors and outdoors in different frequency bands, such as television, AM/FM radio, Bluetooth, ultra-wideband (UWB), Wi-Fi and cellular. Of these, Wi-Fi and cellular signals are most commonly available and extensively deployed in most consumer devices.
In \cite{LTEIO}, an ensemble learning framework was introduced for I/O classification which considered a specific urban setting, including five malls, with data taken using a custom Android application.
However, this work considered signals from only one technology (4G) operating on one frequency band (2.1 GHz), significantly limiting its broader application scope.

In a more recent study \cite{mainn}, authors incorporated both Wi-Fi and cellular data across diverse frequency ranges for classification of I/O environments. Various ML methods were evaluated, including Decision Tree (DT) and Random Forest (RFo), achieving an impressive 99 \% accuracy in environment detection. However, there are certain limitations in this work: \textit{\textbf{(1)}} The two-classes I/O classification falls short in the detection of client devices operating indoors near windows. These devices can transmit at power levels that could cause interference to outdoor incumbents. \textit{\textbf{(2)}} The dataset is unbalanced, comprised predominantly of outdoor-labeled data, resulting in classification errors that are biased towards the outdoor category. Additionally, outdoor data were primarily collected via driving; outdoor 6 GHz APs and clients are more likely to be used while walking rather than driving. \textit{\textbf{(3)}} The proposed ML models exhibited degraded performance when tested with datasets from a new environment, indicating an inability to maintain the same level of accuracy when exposed to unknown locations.

\begin{figure}
    \centering
    \includegraphics[width=8cm,height=3.25cm]{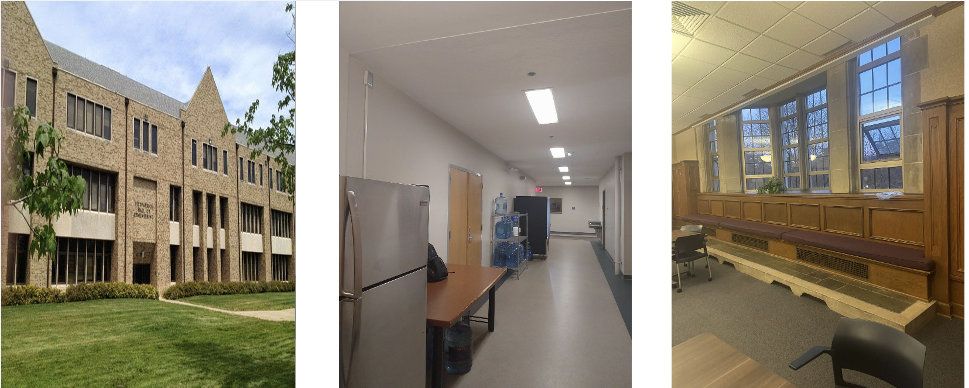}
    \vspace{-1em}
    \caption{Example locations representing outdoor, indoor interior, and indoor near window environments.}
    \label{locs}
    \vspace{-1.5em}
\end{figure}

\subsection{Motivation and Main Contributions}

In this paper, we endeavor to tackle some of the above challenges associated with the classification of I/O status of wireless devices.
The main contributions of this paper are: 


\begin{itemize} [leftmargin=*]

\item \textbf{Comprehensive, Labeled Dataset (\S\ref{sec:sigcap}, \S\ref{sec:preprocessing}):} Our data, collected while walking on the University of Notre Dame (ND) campus, reflects real-world client mobility. We also enhance our models by incorporating SIM operator labels and measurements from newly deployed 6 GHz Wi-Fi.

\item \textbf{Robust Deep Neural Network (DNN) Model (\S\ref{sec:models}, \S\ref{sec:4feat}):} We develop a DNN model that utilizes Wi-Fi, cellular, and GPS data collected through an Android application to predict client environments. This model outperforms traditional ML methods. 

\item \textbf{Refined Labeling (\S\ref{sec:methodology}, \S\ref{sec:4feat}):} We introduce three distinct labels: Outdoor (O), Indoor Near Window (INW), and  Indoor Interior (II) as shown in Fig.~\ref{locs}. This granularity helps us identify indoor devices with greater potential for signal leakage and interference. 

\item \textbf{Extended Windowing Techniques for Accuracy Improvements (\S\ref{subsec:mv}):} We combine multiple data points over an extended period of time and use two techniques to improve accuracy: data aggregation and majority voting. These strategies enhanced the accuracy from typically above 95\% to a perfect classification (100\%).

\item \textbf{Testing on Novel Data (\S\ref{sec:test_novel}):} We validate our models' ability to generalize to unfamiliar locations by testing it on data from new environments. We find that the ML models yield acceptable results in those new locations, without any additional training. We also demonstrate an accuracy increase when the novel dataset is included in the training.

\item \textbf{Open Wireless Dataset for I/O Inference (Table~\ref{tab:ml_features}, \S\ref{sec:methodology}):} We believe the data collected in this work is the first data collection with three distinct labels (O/II/INW) and comprehensive wireless features as shown in Table~\ref{tab:ml_features}.
We provide our dataset for usage beyond the paper's scope at: 
\begin{center}
\textsf{\textbf{\url{https://github.com/hosseinnasi/Environmental-Classification_ML}}}
\end{center}

\end{itemize}

\begin{table} 
\caption{Information in a single record of SigCap data.}
\label{tab:table1}
\vspace{-1em}
 \renewcommand{\arraystretch}{1.15}
\begin{tabular}{ |p{1.5cm}|p{6cm}|  }
\hline
\textbf{Category}& \textbf{Features} \\
\hline
\hline
4G LTE & Physical Cell ID (PCI), Frequency, Bandwidth, Band Number, Reference Signal Received Power (RSRP), Reference Signal Received Quality (RSRQ), Received Signal Strength Indicator (RSSI) \\
\hline
5G & PCI, Frequency, RSRP, RSRQ, Signal to Interference and Noise Ration (SINR) \\
\hline
2.4/5/6 GHz Wi-Fi & Basic Service Set Identifier (BSSID), Frequency, Bandwidth, RSSI  \\
\hline
GPS    &Longitude, Latitude, Altitude, Accuracy  \\
\hline
Time & Time, Date  \\
\hline

Others & Operating SIM, Android Version  \\
\hline

\end{tabular}
\vspace{-1.5em}
\end{table}

\section{Data Collection and ML Methodology}
\label{sec:methodology}

\subsection{Measurement Tools and Locations}\label{sec:sigcap}
We employed client devices running SigCap to collect data across various environments. SigCap is an Android application that passively collects data in every 5 seconds. It records multiple signal features as shown in Table \ref{tab:table1}. These features undergo processing and cleaning before being used in our model to train the network for the I/O classification task.

To mitigate any reliance on external factors like smartphone brands, our data collection used a range of devices from popular brands: Google Pixel 5, Google Pixel 6, Samsung S21, Samsung S22, and Samsung A23. Each device was equipped with a Subscriber Identification Module (SIM) from one of the three operators: AT\&T, Verizon, and T-Mobile.  

\begin{figure}[t]
    \centering
    \includegraphics[width=.9\linewidth]{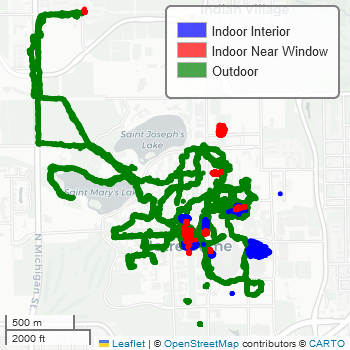}
    \caption{Map of data locations.}
    \label{map}
    \vspace{-1.5em}
\end{figure}

Our dataset comprises records collected in South Bend within the ND campus.
The data was used for both training and testing the ML models. Figure \ref{map} displays the locations where measurement campaigns were conducted. Figure \ref{locs} additionally showcases examples of locations that represent each of the three target environments (O/INW/II). To compare the performance of three-classes classification with the prior two-classes I/O classification, we reuse II and INW data as Indoor (I) data.
Additionally, we sought to evaluate how the ML models, trained on ND data, would perform on datasets from entirely new environments. Therefore, we collected data from Washington DC (DC) using a similar methodology and compare the testing accuracy of our model trained with and without the DC data in \S\ref{sec:test_novel}.

\subsection{Data Processing and Feature Extraction}
\label{sec:preprocessing}

SigCap records data in JavaScript Object Notation (JSON) files, which can be converted into Comma Separated Values (CSV). Each record encompasses various details regarding Wi-Fi, cellular, and GPS information, corresponding to a unique timestamp. Table \ref{tab:table1} provides information about the properties measured within these records. 
The pre-processing procedures are described as follows:

\textbf{(1) Cleaning the dataset}: Entries with invalid values (e.g., incorrect GPS information or values outside the practical ranges for RSRP, RSRQ, RSSI) were removed. Rows lacking channel information (``NaN'' values for Wi-Fi and cellular data frequencies) were also excluded.

\textbf{(2) Grouping Wi-Fi and cellular data by frequency}: To preserve essential frequency band information crucial for classification, we divided all cellular data into two groups based on their operating frequencies: low-band (< 1 GHz) and mid-band (1-10 GHz). Similarly, Wi-Fi data was categorized into 2.4 GHz (2.4-2.5 GHz), 5 GHz (5.15-5.875 GHz), and 6 GHz (5.925-7.125 GHz) bands.

\textbf{(3) Extracting statistical features}: The raw CSV data file, approximately 12 GB in size, is impractical for direct input into our ML models.
Additionally, each column has a varying number of entries due to connections to different numbers of Wi-Fi APs or cellular cells at different times.
Therefore, we extract statistical features (averages, minimums, maximums, and standard deviations) for the RSRP, RSRQ, and frequency features in cellular bands (low- and mid-band), as well as RSSI and frequency features in Wi-Fi data (2.4 GHz, 5 GHz, and 6 GHz).

\textbf{(4) Incorporating SIM Operator Information}: To include the SIM operator information, we assigned a unique number to each of the three major carriers in the USA (Verizon, AT\&T, and T-Mobile). This means that, for now, we are only considering these operators and not others that may, for example, private networks.

\textbf{(5) Normalizing}: Given the diverse range of features, the value ranges for each column can vary significantly. For example, RSRP values for cellular data range from -140 to -20 dBm, while the number of unique LTE cells is a non-negative integer. These raw, non-normalized values could impede ML model convergence. Therefore, we employed MinMaxNormalization to scale all values in each column from 0 to 1.

In summary, Table \ref{tab:ml_features} shows the list of features contained in a single record, while Table \ref{tab:breakdown} shows the amount of data we collected for each label in ND and DC in terms of minutes. Note that the ND dataset is balanced between the labels to prevent over-fitting. As we reuse the II- and INW-labelled data for I data in two-classes classification, we randomly sampled them to keep the amount balanced with the O data.

\begin{table} 
\caption{ML input features.}
\label{tab:ml_features}
\vspace{-1em}
 \renewcommand{\arraystretch}{1.15}
\begin{tabular}{ |p{1.5cm}|p{6cm}|  }
\hline
\textbf{Category}& \textbf{Features} \\
\hline
\hline
4G LTE & Min, Max, Average, and Std. Dev. of RSSI, RSRP, RSRQ for both low- and mid-range frequencies, \# of unique PCIs \\
\hline
5G & Min, Max, Average, and Std. Dev. of RSRP, RSRQ, SINR for both low- and mid-range frequencies, \# of unique PCIs \\
\hline
2.4/5/6 GHz Wi-Fi &Min, Max, Average, and Std. Dev. of RSSI, \# of unique BSSIDs  \\
\hline
GPS and Others    & Horizontal and vertical accuracy, SIM operator\\
\hline
\end{tabular}
\vspace{-1em}
\end{table}

\begin{table} 
\caption{Cumulative amount of data in minutes with percentage of total.}
\label{tab:breakdown}
\vspace{-1em}
 \renewcommand{\arraystretch}{1.15}
\begin{tabular}{ |C{1.5cm}|C{2cm}|C{3cm}| }
\hline
\textbf{Location} & \textbf{Label} & \textbf{Minutes of data} \\
\hline
& II & 1900.4 (33.6\%) \\
\cline{2-3}
\textbf{ND} & INW & 1747.8 (30.9\%) \\
\cline{2-3}
& O & 2014.9 (35.6\%) \\
\hline
\textbf{DC} & II & 1004.2 (71.5\%) \\
\cline{2-3}
\textbf{(novel} & INW & 85.3 (6.1\%) \\
\cline{2-3}
\textbf{dataset)} & O & 313.7 (23.4\%) \\
\hline
\end{tabular}
\vspace{-1em}
\end{table}

\textcolor{black}{The DC dataset is unbalanced due to weather conditions limiting the time available for outdoor measurements. DC INW dataset has fewer samples compared to the other two environments because of fewer windoew.}


\subsection{Model Selection}
\label{sec:models}
To assess the environment classification, we have chosen the following ML models:
    
\textbf{(1) Decision Tree (DT):} This model partitions the data into two or more homogeneous sets by leveraging the most influential features to create distinct groups as effectively as possible. \textcolor{black}{The hyperparameters for training include a maximum depth of 10 and a minimum of 10 samples per leaf.}

\textbf{(2) Random Forest (RFo):} An ensemble variant of the Decision Tree, it employs a majority voting mechanism by aggregating classifications from multiple decision trees, each acting on distinct features. \textcolor{black}{ Similar to DT, we used the same settings for maximum depth (10) and minimum samples per leaf (10), with the addition of 5 estimators (trees).}

\textbf{(3) Deep Neural Network (DNN):} A complex network of interconnected nodes, resembling neurons in the human brain, arranged in multiple layers. The network learns hierarchical representations of data, extracting intricate patterns for classification or regression tasks. 
We trained a DNN model with 4 hidden layers of neurons, containing 64, 32, 16, and 8 neurons, respectively, from the first to the deepest layer. The output layer, depending on the classification type (2- or 3-classes), can have 2 or 3 neurons. The configuration used the Adam optimizer with a learning rate of 0.001. For the final layer, we employed the Softmax activation function.

\textcolor{black}{The hyperparameter selection for the three ML models was conducted through an optimization process, aiming to maximize the accuracy of each model on the testing dataset while minimizing complexity.}
For the testing and training phase using the ND dataset, we allocated 20\% of the entire record for testing the ML models. Additionally, for the DNN model, another 20\% of the dataset was set aside for validation purposes.

\section{Results \& Discussion}

\subsection{Feature Analysis}

\begin{figure}
    \centering
    \begin{subfigure}{0.23\textwidth}
    \includegraphics[width=\linewidth]{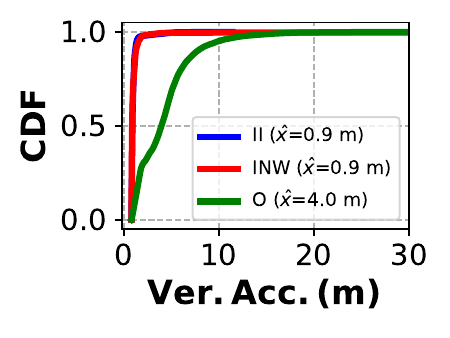}
    \vspace{-2em}
    \caption{Vertical acc.}
    \label{ver_acc}
    \end{subfigure}
    \hfill
    \begin{subfigure}{0.23\textwidth}
    \includegraphics[width=\linewidth]{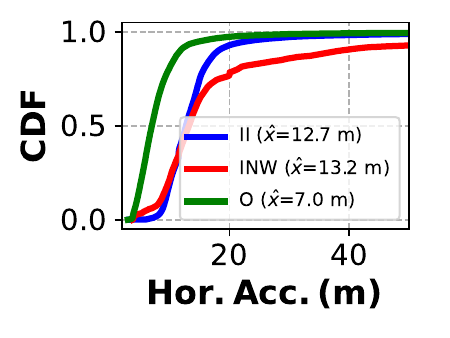}
    \vspace{-2em}
    \caption{Horizontal acc.}
    \label{hor_acc}
    \end{subfigure}
    \\
    \begin{subfigure}{0.23\textwidth}
    \includegraphics[width=\linewidth]{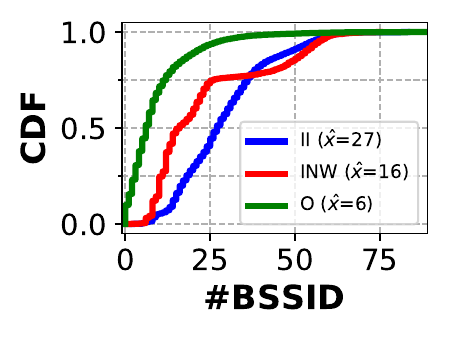} 
    \vspace{-2em}
    \caption{\# of unique 5 GHz Wi-Fi BSSID}
    \label{num_wifi5}
    \end{subfigure}
    \hfill
    \begin{subfigure}{0.23\textwidth}
    \includegraphics[width=\linewidth]{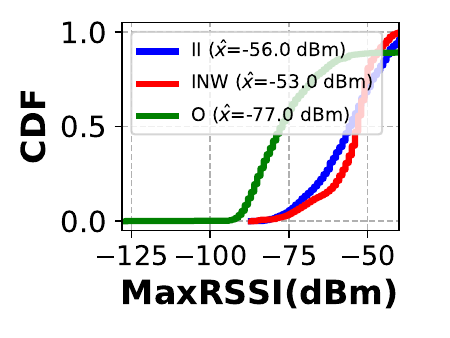}
    \vspace{-2em}
    \caption{Maximum RSSI of 5 GHz Wi-Fi}
    \label{max_wifi5}
    \end{subfigure}
    \\
    \begin{subfigure}{0.23\textwidth}
    \includegraphics[width=\linewidth]{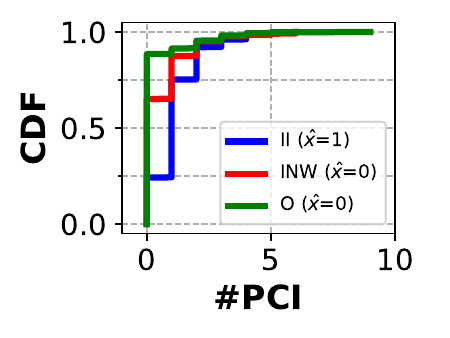} 
    \vspace{-2em}
    \caption{\# of unique low-band LTE PCI}
    \label{lte_pci_low}
    \end{subfigure}
    \hfill
    \begin{subfigure}{0.23\textwidth}
    \includegraphics[width=\linewidth]{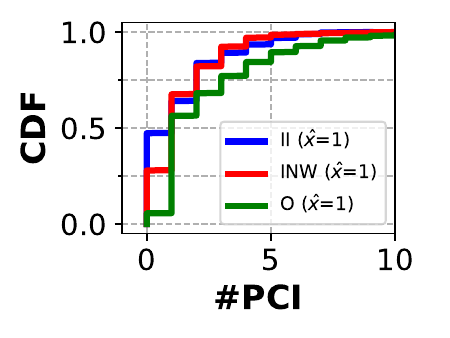}
    \vspace{-2em}
    \caption{\# of unique mid-band LTE PCI}
    \label{lte_pci_mid}
    \end{subfigure}
    \vspace{-1em}
    \caption{CDF plots of input features.}
    \label{fig:foobar}
    \vspace{-1.5em}
\end{figure}

To understand how the extracted wireless features contribute to the classification of II, INW, and O environments, we first illustrate how these features differ across each environment and what enables an ML model to achieve high accuracy. In Figure \ref{fig:foobar}, we present the CDFs of vertical and horizontal Accuracy (defined as the radius in meter of which the true GPS coordinate is contained within 68\% confidence), number of unique BSSIDs in 5 GHz Wi-Fi, maximum RSSI of 5 GHz Wi-Fi, number of PCIs in low-band LTE and number of PCIs in mid-band LTE, respectively on the three environments. As expected, both the number of unique BSSIDs and maximum RSSI of 5 GHz Wi-Fi are higher indoors. We can also observe variations in vertical and horizontal accuracy: vertical accuracy tends to be lower outdoors, whereas horizontal accuracy is generally lower indoors. Conversely, there are no distinction for the number of unique PCI in low- and mid-band LTE: possibly due to the specific deployment of LTE in ND area. Therefore, we present the prior four features as the most important features contribution to the results and we will discuss it in detail in \S\ref{sec:4feat}.

\subsection{Performance of ML Models} \label{sec:4feat}

In this section, we evaluate the performance of the aforementioned ML models in both 2-classes and 3-classes classification tasks. Additionally, we provide performance results for each model when some features are excluded. The following feature sets were considered:

\begin{enumerate}
    \item All features used for the decision-making.
    \item Excluding 6 GHz Wi-Fi.
    \item Excluding 6 GHz Wi-Fi and 5G NR.
    \item Using only the 4 most important features: vertical acc., horizontal acc., \# of unique 5 GHz Wi-Fi BSSID, max. RSSI of 5 GHz Wi-Fi.
\end{enumerate}

\textcolor{black}{Overall, we have 72, 67, 40, and 4 features for feature sets (1), (2), (3), and (4), respectively.}

The rationale behind configuration (2) and (3) is that 6 GHz Wi-Fi and 5G NR are not widely deployed yet. Having appropriate models that account for these variations could be beneficial in improving the accuracy of environmental classification. Moreover, we select 4 most important features to assess the classification accuracy when very few pieces of information are provided to our models. 
These most important features are identified using the SHAP and Scikit-learn libraries in Python.




\subsubsection{Two-Classes Classification}
\label{sec:two_class}

\begin{figure}[h!]
\vspace{-1em}
\centering
\includegraphics[width=.9\linewidth]{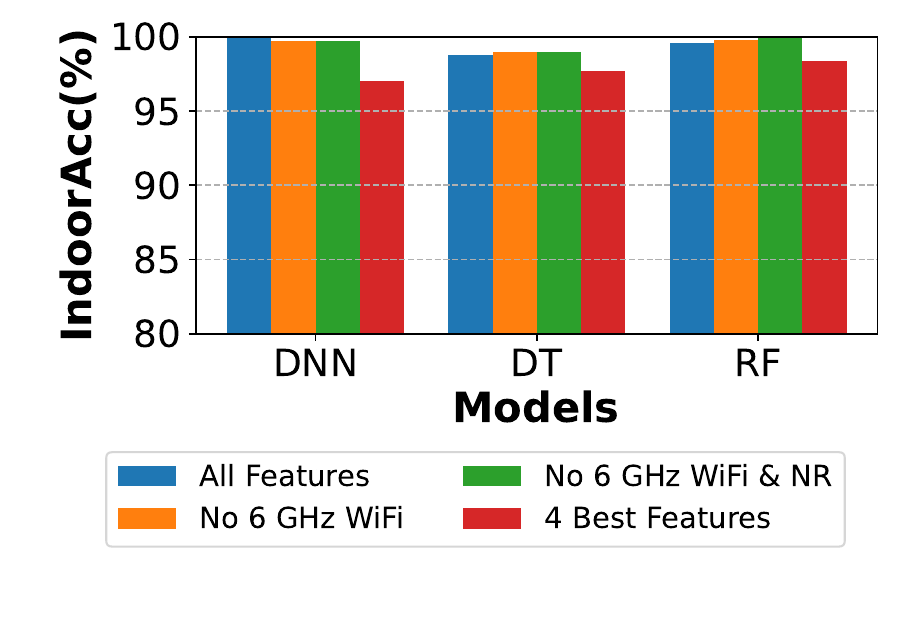}
\vspace{-3em}
\caption{Indoor accuracy by model and feature set, two-classes classification.}
\label{fig:indoor_accuracy_barchart_two_class}
\vspace{-1.5em}
\end{figure}

\begin{figure}[h!]
\centering
\includegraphics[width=.9\linewidth]{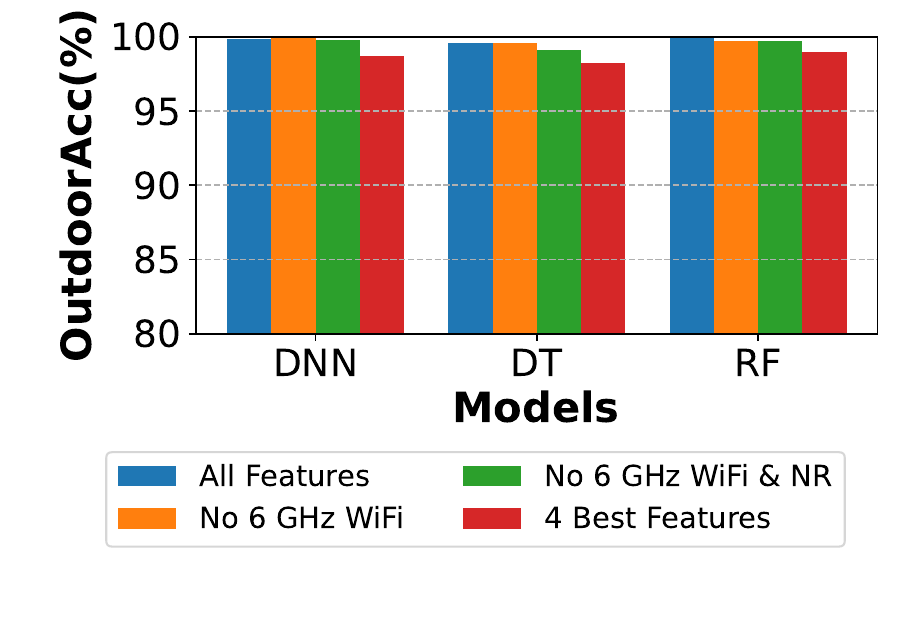}
\vspace{-3em}
\caption{Outdoor accuracy by model and feature set, two-classes classification.}
\label{fig:outdoor_accuracy_barchart_two_class}
\vspace{-1em}
\end{figure}

Figures \ref{fig:indoor_accuracy_barchart_two_class} and \ref{fig:outdoor_accuracy_barchart_two_class}
illustrates the indoor and outdoor accuracy for each feature setup previously discussed in two bar plots. Generally, including all features in the classification task leads to higher accuracy. However, this is not always the case. For example, if Sigcap is run on an indoor device that cannot capture Wi-Fi 6 GHz information and a model using all features is applied, the model might incorrectly infer that the device is outdoors due to the absence of Wi-Fi 6 data. Similarly, a device operating in an environment without 5G NR and Wi-Fi 6 GHz might also produce misleading results when using a model with all features included. As a result, selecting the model with the appropriate feature setup can improve accuracy.

\subsubsection{Three-Class Classification}
\label{sec:three_class}

\begin{figure}
\centering
\includegraphics[width=.9\linewidth]{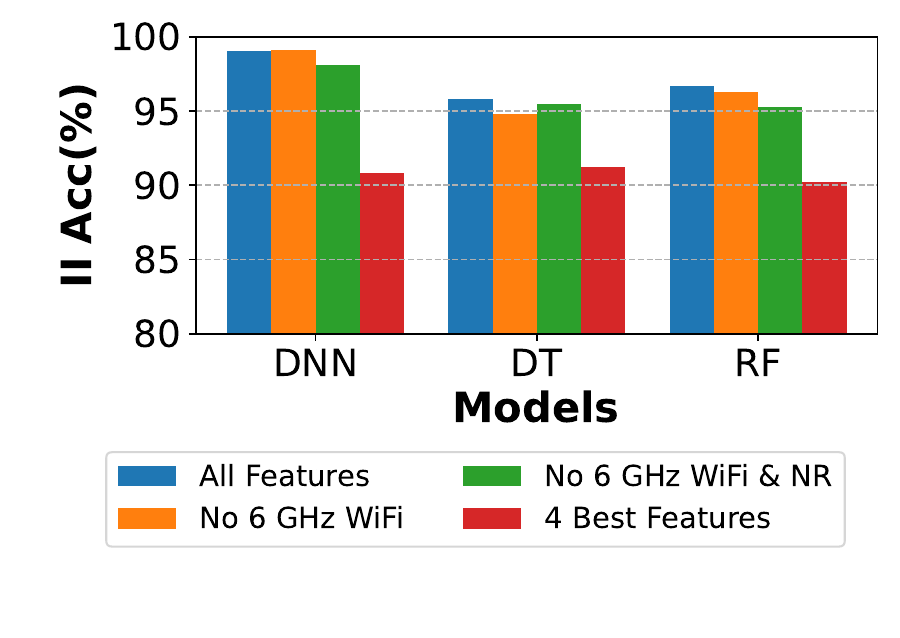}
\vspace{-3em}
\caption{Indoor Interior accuracy by model and feature set, three-classes classification.}
\label{fig:ii_accuracy_barchart_three_class_corrected}
\vspace{-1em}
\end{figure}

\begin{figure}
\centering
\includegraphics[width=.9\linewidth]{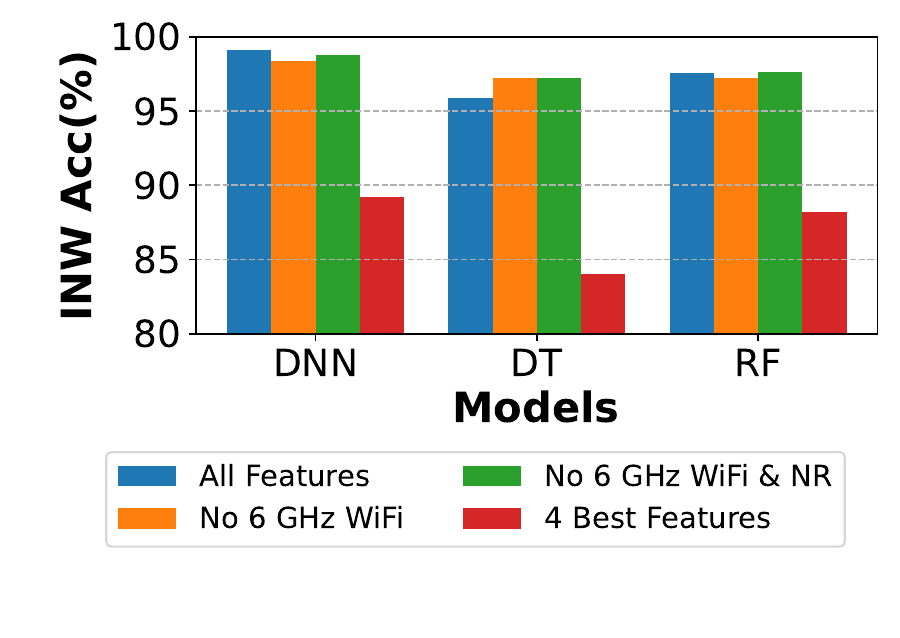}
\vspace{-3em}
\caption{Indoor Near Window accuracy by model and feature set, three-classes classification.}
\label{fig:inw_accuracy_barchart_three_class_corrected}
\vspace{-1em}
\end{figure}

\begin{figure}
\centering
\includegraphics[width=.9\linewidth]{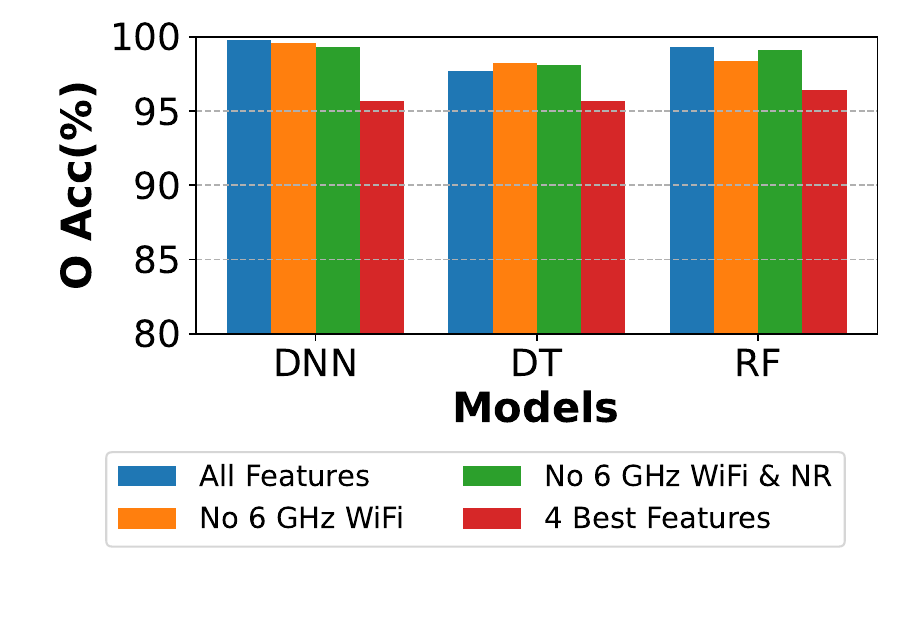}
\vspace{-3em}
\caption{Outdoor accuracy by model and feature set, three-classes classification.}
\label{fig:o_accuracy_barchart_three_class_corrected}
\vspace{-1em}
\end{figure}

The performance of the ML models in the three-class classification task can also be studied similarly to the two-classes classification. Figures \ref{fig:ii_accuracy_barchart_three_class_corrected}, \ref{fig:inw_accuracy_barchart_three_class_corrected}, and \ref{fig:o_accuracy_barchart_three_class_corrected} demonstrate the accuracy results for each model, each environment, and all four possible feature setups. DNN achieved the highest accuracy among all three models. Outdoor accuracy (O) was consistently higher than Indoor Interior (II) and Indoor Near Window (INW) accuracy. For DT and RFo, II and INW accuracy increased when Wi-Fi 6 information and subsequently NR information were removed. However, for DNN, the accuracy for all classes dropped when features were removed. 

\begin{figure}
\centering
\includegraphics[width=.9\linewidth]{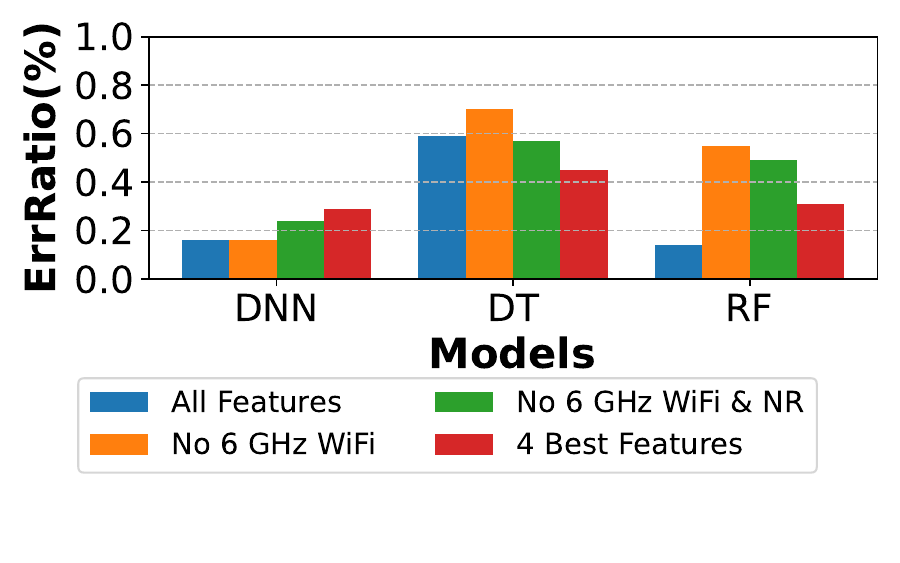}
\vspace{-3em}
\caption{Outdoor to Indoor Interior error rate by model and feature set, three-classes classification.}
\label{3_class_er}
\vspace{-1.1em}
\end{figure}

While overall accuracy is an important metric for evaluation, reducing errors that pose the highest risk of interference is also crucial. For instance, classifying an outdoor device as indoor might lead to the device being deemed safe for transmitting at higher power levels. This could result in interference with incumbent outdoor devices, as the device was actually located outdoors. Therefore, when comparing the performance of each model, in addition to the accuracy rate for each environment, we have also provided the error rate for outdoor-to-indoor interior misclassification.
Figure \ref{3_class_er} illustrates the performance of each ML models for O to II error rate. It is evident that, DNN has the lowest error rate across all the feature setups.

\subsection{Extended Windowing Techniques}
\label{subsec:mv}

As mentioned in Section \ref{sec:sigcap}, SigCap collects wireless data every 5 seconds. However, in real-life scenarios, it is not necessary to make environment predictions every 5 seconds because the device's environment typically remains stable over short periods. Therefore, we introduce Extended Windowing (EW) techniques which leverage all data samples within a certain time window to produce more robust predictions by minimizing the impact of noise. This can be accomplished in two ways:

\begin{enumerate}[leftmargin=*]
    \item \textbf{Data Aggregation (DA)}: This approach involves aggregating data samples over a longer period and then making a decision based on the combined data. Therefore, we examine consecutive samples and accordingly adjust the features. For example, what was once the maximum RSSI for a single row now becomes the maximum RSSI for several consecutive rows. 
    
    \item \textbf{Majority Voting (MV)}: We utilize all the samples in the EW by making individual predictions for each record within an interval and then determine the final prediction based on the most frequently occurring result. This majority voting approach ensures that the most consistent prediction is selected for the entire interval.
\end{enumerate}


The environment predictions should be made after a period of time longer than the SigCap data collection rate (which is 5 seconds).
We chose 30-second intervals as they are long enough to detect environmental changes yet short enough to permit timely adjustments, such as modifying the transmit power limit for Wi-Fi 6E users to avoid interference with outdoor incumbents, as discussed in Section \ref{sec:imp_of_class}.
Therefore, the MV method will take the majority decision from 6 individual predictions, while the DA method makes a decision from a single record created from a combination of 6 consecutive records.


\begin{table}[h!]
\centering
\caption{Accuracy results for three-classes classification with MV technique.}
\label{tab:accuracy_results_three_class}
\vspace{-1em}
\small
\begin{tabular}{|C{1.3cm}|C{1.15cm}|C{1.35cm}|C{1.35cm}|C{1.35cm}|}
\hline
\textbf{Feat. Sets} & \textbf{Model} & \textbf{O ACC(\%)} & \textbf{II ACC(\%)} & \textbf{INW ACC(\%)} \\ \hline
\multirow{3}{*}{\shortstack{All\\Features}}  
    & DNN & 100 & 100 & 100 \\ \cline{2-5}
    & DT  & 100 & 100 & 100 \\ \cline{2-5}
    & RFo  & 100 & 100 & 100 \\ \hline
\multirow{3}{*}{\shortstack{No 6 GHz\\Wi-Fi}} 
    & DNN & 100 & 100 & 100 \\ \cline{2-5}
    & DT  & 100 & 100 & 100 \\ \cline{2-5}
    & RFo  & 100 & 100 & 100 \\ \hline
No 6 GHz  
    & DNN & 100 & 100 & 100 \\ \cline{2-5}
Wi-Fi \&& DT  & 100 & 100 & 99.8 \\ \cline{2-5}
NR & RFo  & 100 & 100 & 99.7 \\ \hline
\multirow{3}{*}{\shortstack{4 Best\\Features}}
    & DNN & 100 & 100 & 100 \\ \cline{2-5}
    & DT  & 100 & 100 & 95.5 \\ \cline{2-5}
    & RFo  & 100 & 100 & 97.5 \\ \hline
\end{tabular}
\vspace{-.5em}
\end{table}

In both two- and three-classes classification, and on all feature set categories, we observe DA technique resulted in accuracy results of around 95\%, while MV shows near-perfect accuracy results. Due to page limitation, we only include Table \ref{tab:accuracy_results_three_class} which shows MV's near-perfect accuracy results for three-classes classification.

\subsection{Testing with Data From a New Environment}
\label{sec:test_novel}

Previously, we evaluated the performance of the ML methods on the testing dataset that came from the same general environment as training, \textit{i.e.}, South Bend. Now, we aim to evaluate the performance of the models when they are being tested with a dataset collected in totally new locations that were not included in the training process. For this purpose, one different set of data was collected in Washington DC. The dataset consists of multiple files, each containing 10 to 20 minutes of measurements, with all samples in a specific file sharing the same label.


For the new dataset, we introduce a new ML method, namely ``Adjustable Classification''. 
This method can be described as follows: For each measurement record, we first select a proper model based on the available feature set. For instance, if one record does not contain 6 GHz Wi-Fi and NR information, we choose the model trained on the respective feature set. The overall accuracy is then calculated as the weighted mean of the collection of records taken in one recording session, with each collection weighted according to the number of records contained.
In Tables \ref{adjustable_2class} and \ref{adjustable_3class}, we compare the overall accuracy of the ML methods on the DC dataset for two and three-classes classification, respectively.

\begin{table}[h!]
\caption{2-classes classification, DC dataset, adjustable classification.}
\label{adjustable_2class}
\vspace{-1em}
\centering
\begin{tabular}{|C{1.3cm}|C{1.15cm}|C{1.35cm}|C{1.35cm}|}
\hline
\textbf{EW Tech.} & \textbf{Model} & \textbf{O ACC(\%)} & \textbf{I ACC(\%)} \\ \hline
\multirow{3}{*}{No EW}  
    & DNN & 89 & 77  \\ \cline{2-4}
    & DT  & 90 & 87  \\ \cline{2-4}
    & RFo  & 99 & 86  \\ \hline
\multirow{3}{*}{MV} 
    & DNN & 91 & 83  \\ \cline{2-4}
    & DT  & 95 & 93  \\ \cline{2-4}
    & RFo  & 100 & 95 \\ \hline
\multirow{3}{*}{DA}  
    & DNN & 87 & 73  \\ \cline{2-4}
    & DT  & 93 & 91  \\ \cline{2-4}
    & RFo  & 99 & 88  \\ \hline
\end{tabular}
\vspace{-1.8em}
\end{table}

\begin{table}[h!]
\caption{3-classes classification, DC dataset, adjustable classification.}
\label{adjustable_3class}
\vspace{-1em}
\centering
\begin{tabular}{|C{1.3cm}|C{1.15cm}|C{1.35cm}|C{1.35cm}|C{1.35cm}|}
\hline
\textbf{EW Tech.} & \textbf{Model} & \textbf{O ACC(\%)} & \textbf{II ACC(\%)} & \textbf{INW ACC(\%)} \\ \hline
\multirow{3}{*}{No EW}  
    & DNN & 89 & 69 & 49 \\ \cline{2-5}
    & DT  & 90 & 77 & 58 \\ \cline{2-5}
    & RFo  & 90 & 91 & 48 \\ \hline
\multirow{3}{*}{MV} 
    & DNN & 91 & 77 & 52 \\ \cline{2-5}
    & DT  & 93 & 78 & 64 \\ \cline{2-5}
    & RFo  & 94 & 93 & 53 \\ \hline
\multirow{3}{*}{DA}  
    & DNN & 87 & 66 & 51 \\ \cline{2-5}
    & DT  & 91 & 84 & 66 \\ \cline{2-5}
    & RFo  & 89 & 86 & 59 \\ \hline
\end{tabular}
\vspace{-1.5em}
\end{table}

\begin{table}[h!]
\caption{3-classes classification, DC and ND dataset combined, all features.}
\label{dc_plus_nd_3class}
\vspace{-1em}
\centering
\begin{tabular}{|C{1.3cm}|C{1.15cm}|C{1.35cm}|C{1.35cm}|C{1.35cm}|}
\hline
\textbf{EW Tech.} & \textbf{Model} & \textbf{O ACC(\%)} & \textbf{II ACC(\%)} & \textbf{INW ACC(\%)} \\ \hline
\multirow{3}{*}{No EW}  
    & DNN & 99.4 & 99.1 & 99.4 \\ \cline{2-5}
    & DT  & 98.1 & 94.7 & 96.4 \\ \cline{2-5}
    & RFo  & 99.0 & 96.1 & 98.0 \\ \hline
\multirow{3}{*}{MV} 
    & DNN & 100 & 100 & 100 \\ \cline{2-5}
    & DT  & 100 & 100 & 100 \\ \cline{2-5}
    & RFo  & 100 & 100 & 100 \\ \hline
\multirow{3}{*}{DA}  
    & DNN & 99.7 & 99.4 & 99.5 \\ \cline{2-5}
    & DT  & 98.0 & 94.9 & 96.9 \\ \cline{2-5}
    & RFo  & 100.0 & 99.9 & 99.9 \\ \hline
\end{tabular}
\vspace{-.5em}
\end{table}

The tables show for both types of classification, accuracy is generally higher in outdoor environments than indoors. This discrepancy is mainly due to the uniformity of outdoor environments across different locations, which results in consistent wireless features. Conversely, indoor environments vary more significantly, affected by the diverse architectures and materials of buildings from one city to another.  

Overall, DT achieves the highest accuracy among the models. Using the MV and DA techniques, it consistently reaches an accuracy of at least 64\% in both classification types and across all environments: O, II, INW.

In comparing MV with DA, the tables clearly show that MV generally outperforms DA across all ML methods in both two and three-class classifications. However, DA yields better results when tested with INW-labeled data.

To address the relatively lower accuracy of the DC testing data compared to the ND data, we combined the DC and ND datasets and repeated the procedures outlined in Section \ref{sec:4feat}, this time using the combined dataset. The results of the three-class classification on the 20\% portion of the testing dataset are presented in Table \ref{dc_plus_nd_3class}. (As the accuracy for the simpler two-class classification task was similarly high, those results are not included here). As shown in the table, the near-perfect accuracy results (mostly above 98\%) demonstrate that as more training data becomes available from diverse environments, both indoors and outdoors, the model performance will improve.
 
\section{Conclusions \& Future Work}

It is clear that the ability to robustly identify indoor and outdoor environments will be important to efficiently manage spectrum in shared spectrum environments. In this paper we demonstrated that such classification using the received RF environment in a smartphone is possible with different ML models, with maximum performance obtained with DNNs. As with any ML methodology, using the right dataset for training is as important as the model itself: towards that end we have collected a comprehensive, labeled, first-of-its-kind dataset for classification, which we continually update with data from different locations: this dataset will be made available publicly. We incorporated "Extended Window" techniques that take advantage of data samples over a longer time period to achieve almost perfect classification accuracy. We have also introduced an adjustable approach to applying ML models to data samples, which can further enhance performance. Finally, we examined the performance of the ML models after incorporating a new dataset into the training set. We observed consistently high accuracy results on the testing dataset, mostly exceeding 98\%. 

Our future work will seek to develop an implementation of the classification models on the smartphone: SigCap data will directly feed into trained models on the phone which will display the environment directly on the phone. \textcolor{black}{Further model refinements will involve leveraging the correlation and time-dependent characteristics of signals by treating the data as a time series. This approach can capture temporal relationships and patterns within the data, potentially enhancing the model's predictive capabilities. Models such as GRU (Gated Recurrent Unit), LSTM (Long Short-Term Memory), and the ROCKET family \cite{ROCKET} of networks are promising for capturing these temporal dynamics, offering deeper insights and improved predictions based on time series data.} 


\section*{Acknowledgements}
This research was funded in part by NSF Grants CNS-2229387, CNS-2346413, and AST-2132700.

\bibliographystyle{ieeetr}
\bibliography{main}

\end{document}